\documentclass[fleqn,12pt,twoside]{article}
\usepackage[headings]{espcrc1}

\usepackage{amsmath}
\usepackage{amssymb}

\readRCS
$Id: espcrc1.tex,v 1.2 2004/02/24 11:22:11 spepping Exp $
\usepackage{graphicx}
\usepackage[figuresright]{rotating}

\providecommand{\openone}{\leavevmode\hbox{\large1\kern-7.3pt\normalsize1}}
\newcommand{\be}{\begin{equation}}
\newcommand{\ee}{\end{equation}}
\newcommand{\ba}{\begin{eqnarray}}
\newcommand{\ea}{\end{eqnarray}}
\newcommand{\rmi}[1]{{\mbox{\tiny #1}}}

\newcommand{\lambdamsbar}{{\Lambda_{\overline {\mbox{\tiny MS}} }}}

\renewcommand{\ln}{{\rm ln}}

\title{The pressure of deconfined QCD for all temperatures and quark chemical potentials}

\author{Andreas Ipp\address[MCSD]{ECT*, Villa Tambosi, Strada delle Tabarelle 286,
I-38050 Villazzano Trento, Italy}}
       
\runtitle{The pressure of deconfined QCD for all temperatures and quark chemical potentials}
\runauthor{Andreas Ipp}

\begin{document}

\maketitle

\begin{abstract}
  A new method for the evaluation of the perturbative
  expansion of the QCD pressure is presented which is valid 
  for all temperatures and quark chemical potentials 
  in the deconfined phase, and worked out up to and including
  order $g^4$.
  This new approach unifies several distinct perturbative approaches
  to the equation of state, and agrees with
  dimensional reduction, HDL and HTL resummation schemes, and the 
  zero-temperature result in their respective ranges of validity.
\end{abstract}

\section{Introduction}

In a recent paper \cite{Ipp:2006ij} Kajantie, Rebhan, Vuorinen, and myself have proposed a new method for calculating the QCD equation of state in a perturbative setup for all temperatures $T$ and quark chemical potentials $\mu$.
Previous perturbative approaches have been limited to certain regions only: when the temperature is large compared to the Debye screening mass $T\gtrsim m_{\rm D}$ (with $m_{\rm D}^2 = g^2 (T^2 + \mu^2 /\pi^2 )$ for $N_c = 3$ colors and $N_f=2$ flavors), hard modes can be integrated out to leave an effective theory for the zero mode, and dimensional reduction applies 
\cite{bn1}-\cite{avpres}. In the opposite case $T\lesssim m_{\rm D}$, a four-dimensional resummation in terms of hard dense loops (HDL) works, giving rise to anomalous contributions in entropy and specific heat of a non-Fermi liquid \cite{Ipp:2003cj}. The latter approach smoothly connects to the well-known zero-temperature limit by Freedman and McLerran \cite{fmcl}.
Our new method covers all of these cases, and provides an independent check on those calculations.

\begin{figure}[htb]
\center{
\includegraphics[width=13.4cm,bb=120 565 445 715]{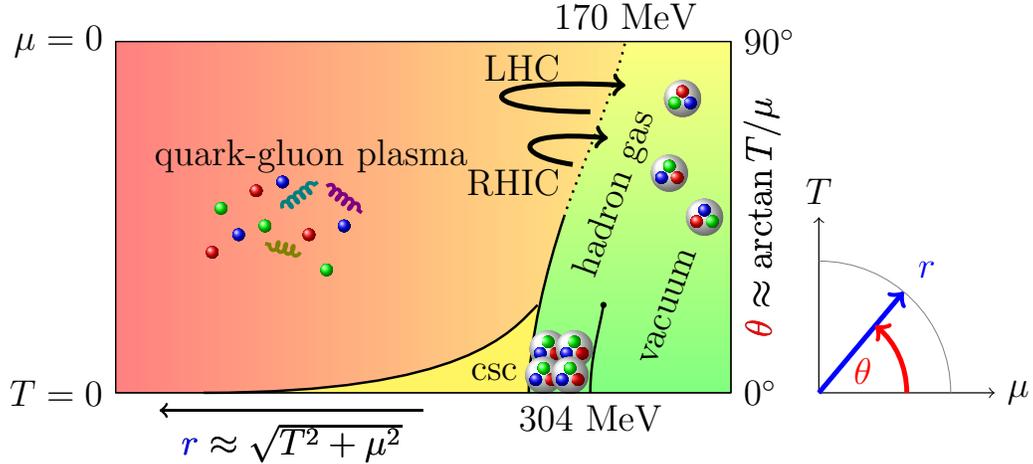}}
\caption{QCD phase diagram in polar coordinates.}
\label{fig:phasediagALL}
\end{figure}

These parametrically distinct regions are suitably visualized in a QCD phase 
diagram in polar coordinates (Figs.~\ref{fig:phasediagALL}-\ref{fig:phasediagPIV}). 
Due to asymptotic freedom, 
weak-coupling techniques apply
at high temperatures or chemical potentials 
$r\approx \sqrt{T^2 + \mu^2}$. We label the abscissa
in Figs.~\ref{fig:phasediagDR} and \ref{fig:phasediagPIV} by 
powers of the coupling $g^2$, $g^3$, \ldots~and the ordinate by parametrically 
different regimes 
$T\sim \mu$, $g \mu$, $g^2 \mu$, 
\ldots~which for larger $r$ 
are displayed as horizontal lines (somewhat deviating from strict polar coordinates).
In the following, we will describe the various existing perturbative approaches,
and connect them by our new method.

\section{Previous methods}

\begin{figure}[htb]
\begin{minipage}[t]{75mm}
\includegraphics[width=7.4cm,bb=115 570 305 705]{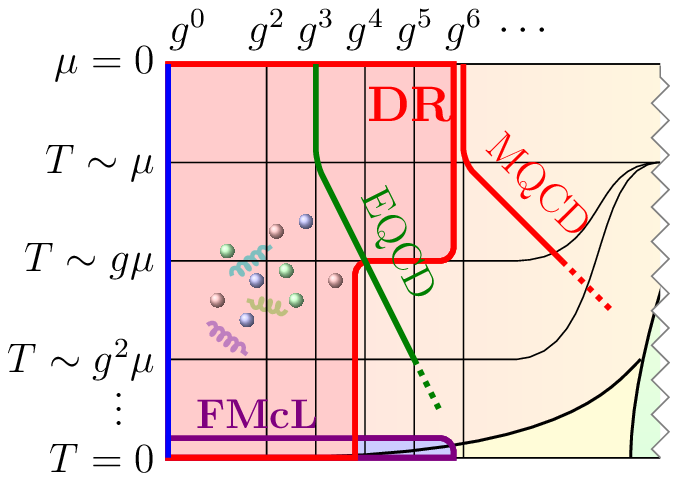}
\caption{Range of validity of dimensional reduction (DR) and the zero-temperature result
(FMcL).}
\label{fig:phasediagDR}
\end{minipage}
\hspace{\fill}
\begin{minipage}[t]{75mm}
\includegraphics[width=7.4cm,bb=115 570 305 705]{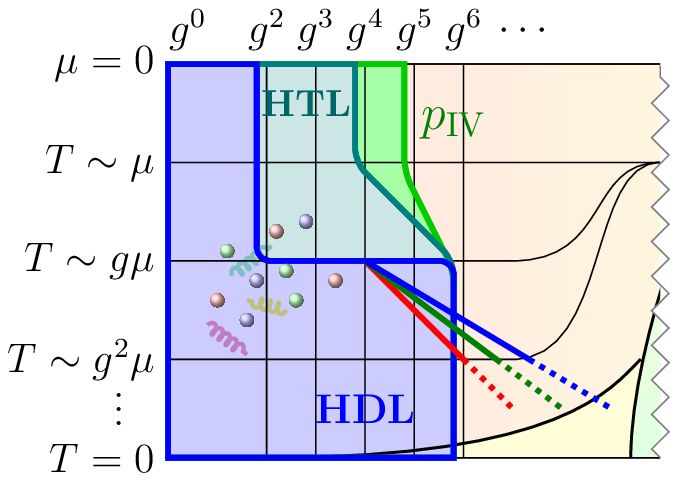}
\caption{Range of validity of HDL and HTL resummations, and the $p_{\rm IV}$ presented here.}
\label{fig:phasediagPIV}
\end{minipage}
\end{figure}

If the temperature $T$ is larger than all other dynamical scales,
one can apply effective field theory methods
\cite{bn1} and 
integrate out the degrees of freedom
corresponding to non-zero Matsubara modes 
to obtain a simpler three-dimensional
(dimensionally reduced)
effective theory, known as 
electrostatic QCD (EQCD).
The effective theory approach can be applied a second time
to remove also the massive longitudinal gluon from the theory,
thus producing an effective three-dimensional pure Yang-Mills theory,
which is called magnetostatic QCD (MQCD).
Practically, the effective Lagrangians 
and effective masses $m_{\rm E}$ and couplings $g_{\rm E}$, $\lambda_E$, $g_M$, \ldots~%
can be obtained by matching
a set of physical quantities between a
generic Lagrangian and the full theory.
The perturbative pressure can then be written as \cite{bn1}
\begin{equation}
p = p_E(T,g) + p_M(m_E^2, g_E, \lambda_E,...)	+ p_G(g_M,..),
\end{equation}
where the first two terms correspond to the coefficients 
of the unit operator of EQCD and MQCD respectively, and the last
term corresponds to the pressure of MQCD, which 
constitutes a fundamentally non-perturbative contribution
to full QCD. Figure \ref{fig:phasediagDR} shows at which orders
the latter two
terms start to contribute, by the lines labeled EQCD and MQCD.
The plot shows for example that the non-perturbative contribution of MQCD 
which sets in at order $g^6$ for $T\gtrsim \mu$, only affects the
pressure at order $g^8$ if the temperature is of the parametric order
$T \sim g \mu$.
So far the perturbative pressure has been obtained 
up to and including order $g^6 \ln g$ \cite{klry} and extended to finite 
chemical potential $\mu$ \cite{avpres}.

The other result depicted in Fig.~\ref{fig:phasediagDR}
is the zero-temperature limit by Freedman and McLerran (FMcL) \cite{fmcl} which is known up to and including 
order $g^4$, with an error of order $g^6 \ln g$.
At order $g^4$ 
dimensional reduction diverges as $T\rightarrow 0$ and does not connect 
to FMcL, 
revealing obviously the breakdown of dimensional reduction.


In the region $T \lesssim g\mu$ perturbation theory requires a different
reorganization which can be most efficiently performed via HDL resummation \cite{Ipp:2003cj} (Fig.~\ref{fig:phasediagPIV}).
A small modification to HTL resummation increases the range of validity of this approach to larger temperatures. The interaction pressure $\delta p \equiv p - p_\rmi{SB} - (p - p_\rmi{SB})|_{T=0}$ ($p_\rmi{SB}$ is the interaction-free Stefan-Boltzmann contribution)
as well as entropy and specific heat
show anomalous logarithmic enhancement of the leading contribution (a signature for non-Fermi-liquid behavior) and fractional powers of the temperature 
in the regime $T \lesssim g\mu$
(for $N_f = 2$, $N_c = 3$)
\begin{equation}
\delta p = \frac{g^2 \mu^2 T^2}{9\pi^2}
			\left\{ \ln  \frac{g \mu}{\pi T} 
			-0.7906 - 2.9401 \left( \frac{\pi T}{ g\mu} \right)^{2/3}
			+2.2312 \left( \frac{\pi T}{ g\mu}\right)^{4/3}
			+\dots \right\} .
			\end{equation}
The fractional powers are indicated by lines starting from $T\sim g\mu$ and $g^4$ in Fig.~\ref{fig:phasediagPIV}, and form the correct continuation of the plasmon term 
$\propto T m_{\rm D}^3$
(EQCD line in Fig.~\ref{fig:phasediagDR}).

\section{The new approach}

\begin{figure}[htb]
\begin{minipage}[t]{70mm}
\begin{center}
\includegraphics[width=70mm]{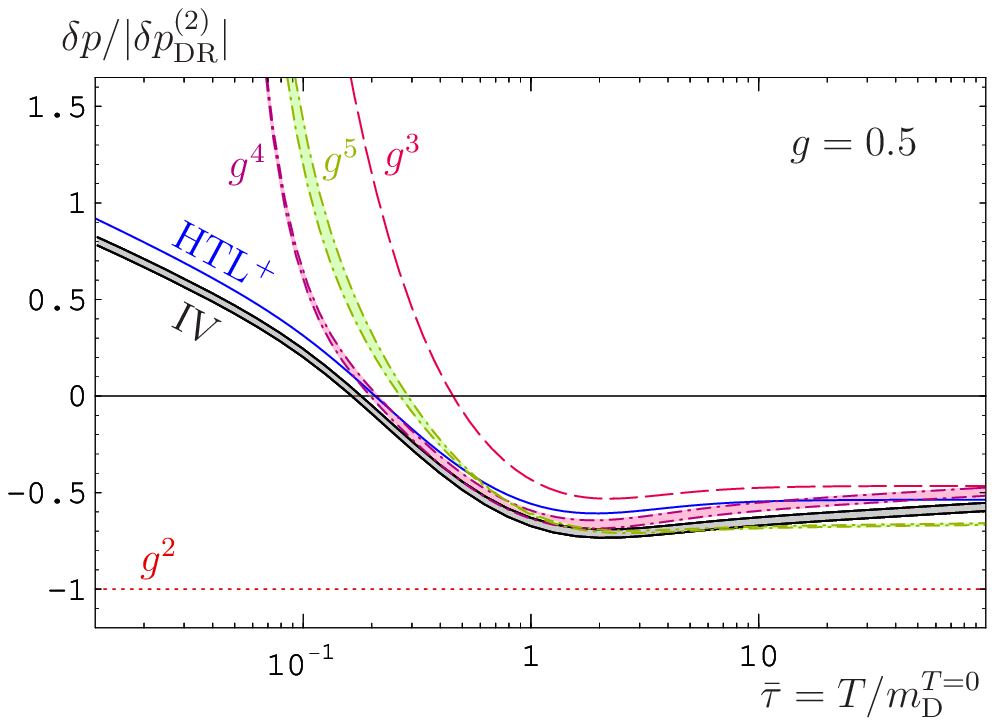}
\caption{
Thermal contribution to the interaction pressure $\delta p$
as a function of $T/m_\rmi{D}^{T=0}$ for fixed chemical potential $\mu$
and coupling $g=0.5$.
When two lines of the same type run close to each other, they differ by
changing the renormalization scale $\lambdamsbar=\mu \,...\, 4\mu$.
\label{fig:p05}}
\end{center}
\end{minipage}
\hspace{\fill}
\begin{minipage}[t]{83mm}
\begin{center}
\includegraphics[width=79mm]{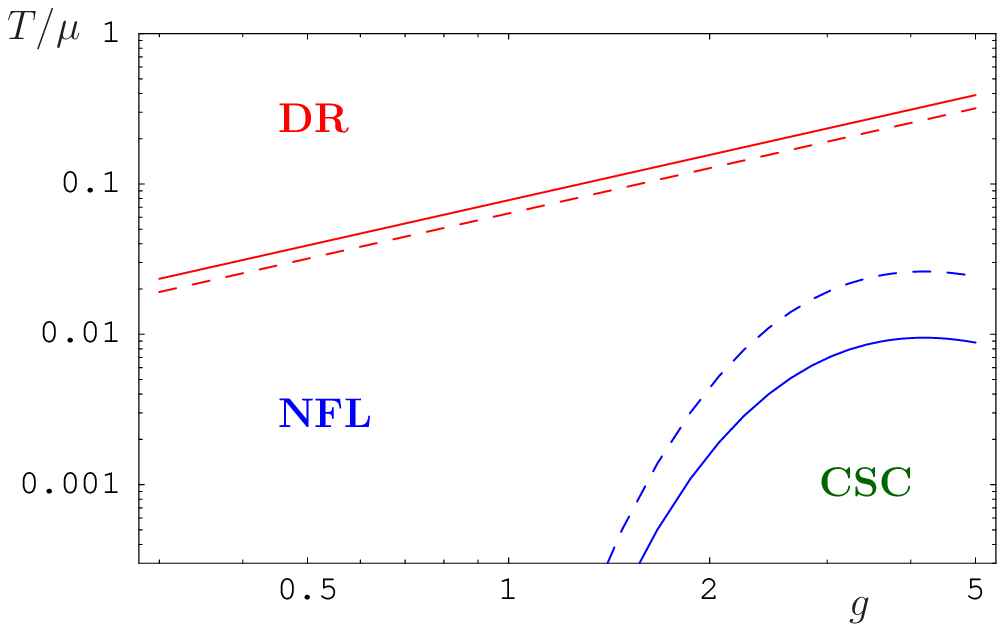}
\caption{
The dividing line between the regime of dimensional reduction (DR)
and that of non-Fermi-liquid behavior (NFL)
for $N_f=3$ (full lines) and $N_f=2$ (dashed
lines), in comparison with the
weak-coupling result
for the
critical temperature of color superconductivity (CSC)
when extrapolated to large coupling.
\label{fig:Tmu}}
\end{center}
\end{minipage}
\end{figure}

Our new approach \cite{Ipp:2006ij} is motivated by the observation 
that only a subset of all diagrams 
that contribute to the
pressure exhibit infrared divergent behavior in the limit $T\rightarrow 0$.
The problematic set consists of the two-gluon reducible diagrams 
(2GR; vacuum bubble diagrams that can
be separated 
by cutting two gluon lines).
Up to order $g^4$, the following classes of 2GR diagrams require resummation:
\vspace{-.2cm}
{\center{\includegraphics[width=16cm]{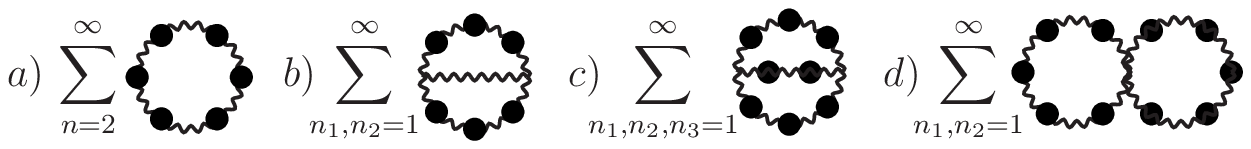}}}
where the black 
dots represent the full one-loop gluon polarization tensor. This is 
only known numerically, 
but the contribution of diagram $(b)$ to the pressure up to $\mathcal{O}(g^4)$ can be expressed analytically, and diagrams $(c)$ and $(d)$ would only contribute beyond that order.
Only the contribution from diagram $(a)$ has to be obtained by means of numerical integration 
(using the methods developed at large $N_f$ \cite{ippreb}).
The pressure through
order $g^4$ (labeled by the roman numeral IV) valid for all $T$ and $\mu$ is finally given by
the sum of 
analytically calculable pieces
and the ring diagrams $(a)$ which have 
to be integrated numerically:
$p_{\rm IV} = p_{\rm anl}+p_{\rm ring}^{\rm safe}+{\mathcal O}(g^5T\mu^3) +
	{\mathcal O}(g^6\mu^4)$.

Figure \ref{fig:p05} shows that dimensional reduction (curves labeled $g^2$, $g^3$, and $g^4$)
ceases to be
applicable when $T\lesssim 0.2 m_{\rm D}$.
At this point, non-Fermi-liquid effects take over that can only be
described by HDL/HTL resummation or the new $p_{\rm IV}$. 
In Fig.~\ref{fig:Tmu}, this separation line $T\approx 0.2 m_{\rm D}$
is plotted along with the weak-coupling result for the critical temperature of color superconductivity~\cite{Brown}.
While at larger couplings significant non-perturbative modifications can be expected,
at weaker couplings
one observes a clear separation of regimes with qualitatively different weak-coupling descriptions.


\end{document}